\documentstyle[epsf]{l-aa}

\def\htre{~$h^{-3}$ Mpc$^3$~}

\begin{document}
\thesaurus{ 11(11.04.1; 11.12.2; 12.03.3; 12.12.1) }
\title{ The ESO Slice Project (ESP) galaxy redshift survey:
\thanks{based on observations collected at the European Southern Observatory, 
La Silla, Chile. 
Table 3 is only available (and Table 2 is also available) in 
electronic form at the CDS via anonymous ftp to cdsarc.u-strasbg.fr 
(130.79.128.5) or via http://cdsweb.u-strasbg.fr/Abstract.html } }
\subtitle{ III. The Sample }
%
\author{
G.Vettolani \inst{1}
\and
E.Zucca \inst{2,1}
\and
R.Merighi \inst{2}
\and
M.Mignoli \inst{2}
\and
D.Proust \inst{3}
\and
G.Zamorani \inst{2,1}
\and
A.Cappi \inst{2}
\and
L.Guzzo \inst{4}
\and
D.Maccagni \inst{5}
\and
M.Ramella \inst{6}
\and
G.M.Stirpe \inst{2}
\and
A.Blanchard \inst{7}
\and
V.Cayatte \inst{3}
\and
C.Collins \inst{8}
\and
H.MacGillivray \inst{9}
\and
S.Maurogordato \inst{10,3}
\and
R.Scaramella \inst{11}
\and
C.Balkowski \inst{3}
\and
G.Chincarini \inst{5,12}
\and
P.Felenbok \inst{3} 
}
%
\institute{ 
Istituto di Radioastronomia del CNR, 
via Gobetti 101, 40129 Bologna, Italy
\and
Osservatorio Astronomico di Bologna, 
via Zamboni 33, 40126 Bologna, Italy
\and
Observatoire de Paris, DAEC, Unit\'e associ\'ee au CNRS, D0173 et \`a
l'Universit\'e Paris 7, 5 Place J.Janssen, 92195 Meudon, France
\and
Osservatorio Astronomico di Brera, 
via Bianchi 46, 22055 Merate (LC), Italy
\and
Istituto di Fisica Cosmica e Tecnologie Relative, 
via Bassini 15, 20133 Milano, Italy
\and
Osservatorio Astronomico di Trieste, 
via Tiepolo 11, 34131 Trieste, Italy
\and
Universit\'e L. Pasteur, Observatoire Astronomique, 
11 rue de l'Universit\'e, 67000 Strasbourg, France
\and
Astrophysics Research Institute, Liverpool John--Moores University, 
Byrom Street, Liverpool L3 3AF, United Kingdom
\and
Royal Observatory Edinburgh, 
Blackford Hill, Edinburgh EH9 3HJ, United Kingdom
\and
CERGA, Observatoire de la C\^ote d'Azur, 
B.P. 229, 06304 Nice Cedex 4, France
\and
Osservatorio Astronomico di Roma, 
via Osservatorio 2, 00040 Monteporzio Catone (RM), Italy
\and
Universit\`a degli Studi di Milano, 
via Celoria 16, 20133 Milano, Italy
}
%
%
\offprints{Elena Zucca (zucca@astbo1.bo.cnr.it)}
\date{Received 00 - 00 - 0000; accepted 00 - 00 - 0000}
\maketitle
\markboth {G.Vettolani et al.: 
The ESP galaxy redshift survey: III. The Sample}{}
\begin{abstract}
The ESO Slice Project (ESP) is a galaxy redshift survey extending over about 
23 square degrees, in a region near the South Galactic Pole. The survey is 
$\sim 85\%$ complete to the limiting magnitude $b_J=19.4$ and consists of 3342 
galaxies with redshift determination.
\\
The ESP survey is intermediate between shallow, wide angle samples and very
deep, one--dimensional pencil beams; the spanned volume is $\sim 5 \times 
10^4$ \htre at the sensitivity peak ($z \sim 0.1$).
\\
In this paper we present the description of the observations and of the data 
reduction, the ESP redshift catalogue and the analysis of the quality of the 
velocity determinations. 
\end{abstract}

\section{Introduction}

The ESO Slice Project (ESP) galaxy redshift survey, which is described 
in Vettolani et al. (1997, hereafter Paper I), extends over a strip of 
$\alpha \times \delta = 22^o \times 1^o$, plus a nearby area of $5^o \times 
1^o$, five degrees west of the main strip, in the South Galactic Pole region. 
The right ascension limits are $ 22^{h} 30^m$ and $ 01^{h} 20^m $, at a mean 
declination of $ -40^o 15'$ (1950). We have covered this region with a regular 
grid of adjacent circular fields, with a
diameter of 32 arcmin each, corresponding to the field of view of the multifiber
spectrograph OPTOPUS (Lund 1986, Avila et al. 1989) at the 3.6m ESO telescope. 
The total solid angle of the spectroscopic survey is 23.2 square degrees.

This paper presents the survey data (photometry, spectroscopy, completeness,
etc.) which are necessary for a comprehensive study of the sample.
It is organized as follows: 
in Section 2 we describe the photometric sample, in Section 3 the observations 
and data reduction and in Section 4 the redshift determination.
In Section 5 we present the catalogue, in Section 6 we discuss the possible
biases in the sample and the velocity errors, and finally Section 7 provides 
a summary.

\section{The photometric sample}

The galaxy catalogue has been extracted from the Edinburgh--Durham Southern
Galaxy catalogue (hereafter EDSGC, Heydon--Dumbleton et al. 1988, 1989) 
which has been obtained from COSMOS (MacGillivray \& Stobie 1984) scans of 
SERC J survey plates.
The EDSGC has a $95 \%$ completeness at $b_J \leq 20.0$ and an estimated 
stellar contamination $\leq 10 \%$ (Heydon-Dumbleton et al. 1989). 
The systematic intra--plate photometric errors are less than $\sim 0.03$ 
magnitudes and the inter--plate limiting magnitude variation is less than
$\sim 0.04$ magnitudes (Heydon-Dumbleton et al. 1988). 

In order to obtain an external estimate of the photometric accuracy of the
sample, we have compared the EDSGC magnitudes with CCD photometry, from
a multicolour survey of a subsample of ESP galaxies
(Garilli et al. in preparation). 
Preliminary analysis of these data, obtained with the 0.9m Dutch/ESO telescope
for about 80 galaxies in the magnitude range $16.5 \leq b_J \leq 19.4$,
shows a linear relation between $b_J$(EDSGC) and $m_B$(CCD), 
with a dispersion ($\sigma_M$) of about 0.2 magnitudes around the fit. 
Since the CCD pointings cover the entire right ascension range of our 
survey, this $\sigma_M$ includes both statistical errors within single
plates and possible plate--to--plate zero point variations.

All EDSGC galaxies in the ESP region have been examined visually on high 
contrast enlarged ($\times 8.4$) reproductions of the ESO--SERC atlas plates. 
A small percentage ($\la 2 \%$) of the catalogue entries was discarded on 
the basis of the visual inspection (portions of star spikes, bright galaxy 
spiral arms ``broken" into multiple entries, etc...). These objects, although 
listed in the original EDSGC catalogue, were not included in the ESP sample
and therefore are not listed in the final catalogue. 

At visual examination 287 objects appeared as formed by two components, not 
separated by the EDSGC deblending algorithm. These 287 objects, which on average
have a magnitude distribution fainter than that of the total sample,
could be either two galaxies almost in contact, or a galaxy
with a nearby star, or a pair of stars, or a galaxy with a prominent HII region.
Being difficult to determine the nature of each object on the basis of the
visual inspection and to estimate the magnitude of each component
separately, we have kept these objects as single entries in the ESP catalogue
in order not to bias {\it a priori} the sample.

The number of objects in the photometric ESP sample is 4487. 
 
\section{Observations and data reduction}

\subsection{Instrumentation}

Spectroscopic observations were performed at the ESO 3.6m telescope in
La Silla. The main body of data was obtained in six observing runs 
with the OPTOPUS multifiber spectrograph at the Cassegrain focus 
(Lund 1986, Avila et al. 1989). Further observations were obtained in 1994 with 
the multifiber spectrograph MEFOS (Avila et al. 1995, Felenbok et al. 1997) 
at the prime focus of the same telescope. A log of the observations is given 
in Table 1.

\begin{table}
\caption[]{OPTOPUS--MEFOS observing log}
\begin{flushleft}
\begin{tabular}{lclc}
\hline\noalign{\smallskip}
 Run & Date & Set-up & \# Obs. Fields \\
\noalign{\smallskip}
\hline\noalign{\smallskip}
 \#~1 & 02--05/09/91 & OPTOPUS+TEK\#16  & 18 \\
 \#~2 & 12--15/10/91 & OPTOPUS+TEK\#16  & 16 \\
 \#~3 & 21--24/09/92 & OPTOPUS+TEK\#16  & 22 \\
 \#~4 & 24--27/10/92 & OPTOPUS+TEK\#16  & 19 \\
 \#~5 & 08--12/09/93 & OPTOPUS+TEK\#32  & 30 \\
 \#~6 & 16--18/10/93 & OPTOPUS+TEK\#32  & 15 \\
 \#~7 & 27--30/10/94 & MEFOS+TEK\#32    & 15 \\
\noalign{\smallskip}
\hline
\end{tabular}
\end{flushleft}
\end{table}
\par
OPTOPUS has a bundle of 50 optical fibers (Polymicro FHP with 320~$\mu$m core
and enhanced blue sensitivity), which are manually plugged into aluminium
plates with holes and fiberholders at the galaxy positions.
These positions are corrected for differential refraction between the
wavelength centre of the autoguider sensor and the centre of the spectral
range, at the expected zenith angle of observation.
The plates are mounted at the Cassegrain focal plane and have a diameter of 
32 arcmin on the sky. The core of the fibers corresponds to a sky aperture of 
2.3 arcsec in diameter. At the opposite end the fibers are coupled to an
\hbox{${\it f}/8$} collimator of the Boller \& Chivens spectrograph,
forming the entrance slit. 
\par 
Once mounted, the plates are centred using four relatively bright stars at
the periphery of the plate, observed through fibers going to an intensified 
camera. Exposure guiding is
accomplished by means of two fiber bundles, which are used to image two
relatively bright stars in the field through holes in the plate.
The mechanical constraints are such that the minimum
object--to--object separation is 24.6 arcsec
and minimum object--to--guide star separation is 64.3 arcsec.
\par
For the present program we used the ESO grating \# 15 (300 lines mm$^{-1}$ 
and blaze angle of $4^{\circ}~18^{\prime}$), allowing a dispersion
of 174~\AA~mm$^{-1}$ with an \hbox{${\it f}/1.9$} blue camera
in the wavelength range 3750--6150~\AA.
Initially the detector was a Tektronix 512~$\times$~512 CCD 
with a pixel size of 27~microns, corresponding to 4.5~\AA/pixel, 
a velocity bin of $\simeq$~270~km/s at 5000~\AA.
After the October 1992 observing run, the detector was replaced with a
new Tektronix 
thinned, back--illuminated CCD (ESO~\#32), which also had a pixel size of 
27~microns, giving a similar spectral coverage but a better blue sensitivity.
\par
The observing time for each field was one hour, split into two 30 mins 
exposures to ease the removal of cosmic rays. Keeping the telescope
in the same position, exposures with a quartz-halogen lamp as well as of a 
Helium comparison lamp were taken immediately before and after the science
exposures. Depending on the length of the night we observed from 4 to 6
different fields per night choosing the sequence of the fields to be observed
in such a way so as to minimize the zenith distance and hence the differential 
refraction. 
\par
Further observations were accomplished in October 1994 with the multifiber 
spectrograph MEFOS, at the prime focus of the 3.6m telescope.
\par
The MEFOS Spectrograph has 30 robotic arms in a ``fishermen-around-the-pond''
configuration over a field of one degree diameter. One arm is used for
guiding purposes, while the other 29 arms carry two spectroscopic fibers
each, one fiber for the object and one for the sky. Because
of the prime focus scale, the optical fibers (Polymicro FBP) have a
135~$\mu$m core, corresponding to a sky aperture of 2.6 arcsec.
The object--sky fiber separation is 60 arcsec, and the minimum
object--to--object distance is 28 arcsec.
Each arm also carries an imaging fiber bundle which is used to accurately 
centre the fibers on the targets.
The output ends of the spectroscopic fibers are arranged in a line so as to
form the entrance slit of the same Boller \& Chivens spectrograph used for
OPTOPUS. With a new optimized \hbox{${\it f}/3.03$} collimator the
spectrograph--CCD configuration was the same as used for the previous OPTOPUS
observations.

\subsection{Data Reduction}

The data reduction was performed using the {\sc apextract} package as
implemented in {\sc iraf}\footnote{{\sc iraf} is distributed by the National
Optical Astronomy Obser\-vatories, which is operated by AURA Inc. for the NSF.}.
For each exposure first we identified and followed the spectra on the white 
lamp frame. Then the solutions obtained for the white lamps were
applied to extract the one--dimensional spectra both in the calibration 
and science frames. Using {\sc apextract} we performed an ``optimal extraction''
(Horne 1986), also detecting and replacing the highly deviant bad pixels and
cosmic ray hits.
\par
The individual wavelength calibration for each fiber is derived from the
corresponding arc spectrum. We used a fourth--order polynomial fit with
typical rms errors of 0.2--0.3~\AA.
The accuracy of the calibration was estimated from the measured wavelengths
of two [OI] sky lines ($\lambda\lambda$~5577,6300), which were
always within $\pm~1$~\AA \ (less than a quarter of pixel) from 
their expected positions.
\par
A critical point in fiber spectroscopy is the sky subtraction because
it is impossible to measure the background locally and through the
same aperture (fiber) of the object (see Parry \& Carrasco 1990
or Wyse \& Gilmore 1992 for a detailed discussion).
We adopted the following strategy. For OPTOPUS observations we dedicated 
at least four apertures to the sky in each field.
We defined predetermined positions on each plate which were verified not to have
any object at the limit of SERC J plates. For MEFOS the sky fibers were
as many as the target objects. With both instruments we performed a
``mean--sky'' subtraction method (Cuby \& Mignoli 1994).
\par
The main problem is the fiber throughput determination, because of the
different transmission of each wave\-guide which could also vary from one
exposure to the other.
The most intense and isolated sky line in our spectral range
is the [OI]5577 emission line, and we used its flux as an estimator
of the fiber trasmittance. We directly measured the line counts
on the calibrated spectra by mean of a Gaussian profile fit, after subtracting
the underlying continuum (due to both sky and object, if any, contribution).
Under the assumption that the intensity of the night spectrum does not vary
appreciably within the telescope field, the flux of this sky line is the best 
estimator of the fiber throughput. In fact it is easily
measurable both in the sky and object spectra and it is subject to the
same temporal transmittance variations as the object flux (e.g. due to the
differential
fiber flexures during the exposure). In order to reveal possible contaminations
(cosmic ray hits or object emission/absorption lines) that could affect
the fiber throughput estimate, we also computed the mean and the
dispersion of the ratio between the peak of the continuum--subtracted counts 
of the [OI]5577 and the same quantity for the [OI]6300 sky line:
spectra with unusual value of this ratio were re--examined to check the
reason for the discrepancy. In all the relatively few cases in which 
this has occurred, 
the problem was overcome by interactively fitting the [OI]5577 line
and eliminating the cosmic ray contaminations,
or was due to an underestimate of the sky [OI]6300 emission
because of the coincident absorption feature in stellar objects.
\par
After the normalization to account for the relative fiber transmission,
a ``mean sky'' was determined using all the sky spectra (four or more for
OPTOPUS, up to 29 for MEFOS) and subtracted from all the ``object plus sky'' 
spectra of the same exposure. This procedure was repeated separately for each 
of the two 30 mins exposures.
The comparison of the two exposures of the same field revealed the
remaining cosmic rays spikes, which were eliminated by interpolating
the adjacent pixels.
\par
The sky--subtraction accuracy is difficult to gauge. The residual counts
in the sky--subtracted sky spectra provide a good estimator of this, but
caution is necessary: indeed, the simple rms value of the sky--subtracted
sky spectrum could overestimate the quality of the sky subtraction because
it does not take account for ``bias residual'' if the average background
has been over/underestimated (see Wyse \& Gilmore 1992). 
Therefore we adopted, as estimator of the sky--subtraction accuracy, the
``average absolute sky--residuals'' quantity

\begin{equation}
 Q_i = \biggl\langle\Bigl\vert{sky_i - \langle sky \rangle_i \over 
{ \langle sky \rangle_i }} \Bigr\vert\biggr\rangle_\lambda. 
\end{equation}

This quantity has been measured, resulting in the range of $2.8\div 7.5$\%
with a typical value of 4\%.

\begin{figure*}
\epsfysize=13cm
\epsfbox{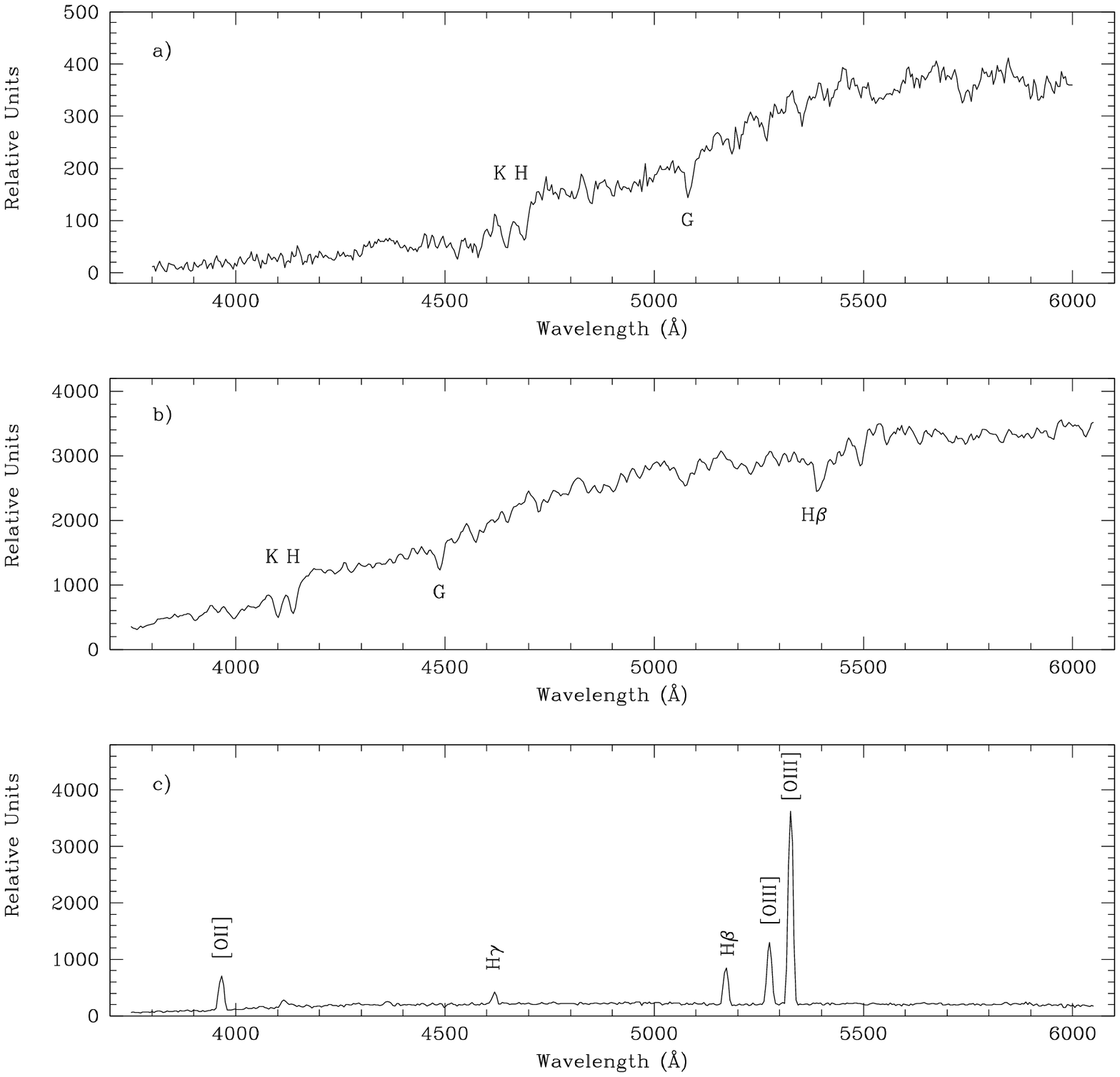}
\caption[]{Three examples of spectra: a) Standard quality spectrum ($R=3.8$)
of a galaxy with $v_{abs}=54315\pm 46$ km/s. b) High quality spectrum 
($R=20.2$) of a galaxy with $v_{abs}=12764\pm 20$ km/s. c) Spectrum with 
prominent emission lines of a galaxy with $v_{emiss}=19110\pm 10$ km/s.
}
\end{figure*}

\section{Redshift determination}

Redshifts were measured for absorption--featured spectra using 
the task {\sc xcsao} (Kurtz et al. 1992), a cross--correlation program 
(Tonry \& Davis 1979) developed at the Smithsonian Astrophysical
Observatory as a part of the {\sc rvsao} package, a contributed {\sc iraf} 
package.
\par
The program allows the user to adapt the several working parameters to the 
characteristics of its own data--set. It is in fact possible to control if and 
how to rebin the data in log wavelength, to set an initial guess for the radial 
velocity, to define the dimension of the apodization region to be applied to 
the spectra, to fit and subtract the continuum from both spectra and templates, 
to choose the appropriate Fourier filtering and how to fit the peak of the 
correlation function.
Once these parameters are defined, the program can be run in batch mode.
An interactive graphical mode is available, during which all the main 
parameters can be modified and tested.
\par
Redshifts for emission line objects were instead determined using {\sc emsao} 
(Mink \& Wyatt 1995), the {\sc xcsao} companion task. {\sc emsao} was 
developed to find emission lines automatically, to compute redshifts 
for each identified line and to combine them into a single radial velocity. 
The results may be graphically displayed or printed. The graphic cursor can be 
used to interactively change fit and display parameters.

\subsection {Templates}

A crucial point for minimizing the error of the cross--correlation is the 
construction of a good template system, i.e. a set of spectra closely 
reproducing the characteristics of the observed spectra, having for example a 
range 
of relative absorption line depths similar to that of the sample galaxies of 
different morphological types.
\par
After the first observing run we realized that some of the spectroscopic stars, 
which were misclassified as galaxies in the EDSGC, could represent such a set 
of templates.
\par
A system of eight stars (6 of F type and 2 of K type) was chosen and used.
Note that the use of such cold stars is a valuable criterion for 
cross--correlation procedure, since the profile of strong lines such as
the calcium $K$ and $H$, the $G$ band, $H\beta$ and $MgI$ is not affected by 
gravity and/or metal
abundance: it remains narrow, with a Doppler core and Stark broadened wings.
\par
We made the assumption that such a group of stars could define a self 
consistent system with an average radial velocity of zero km/s. This assumption 
has been verified cross--correlating against this template system three 
standard radial velocity stars we purposely observed with OPTOPUS. 
We measured a zero point shift of $-7.4 \pm 3.8$ km/s; this correction has not
been applied to the data of the catalogue. 

\subsection{Redshift measurement}

Absorption spectra were measured with {\sc xcsao} and we decided to adopt as 
the ``absorption velocity" the one associated with the minimum error in km/s 
from the cross--correlation against the eight stellar templates. 
In the great majority of cases, this coincided also with the maximum R 
parameter of Tonry \& Davis (1979). 
Generally, the best performing templates were the F stars, but in a few cases 
better measurements were obtained from the two K templates. 
\par
Spectra showing both absorption and emission features were generally measured 
with both tasks ({\sc xcsao} and {\sc emsao}). Emission features were manually 
excised prior to performing the cross--correlation analysis. 
\par
As a consistency check, all spectra were examined also visually, in order
to verify the assigned velocity.
A few examples of spectra of different quality are shown in Figure 1. 

\begin{table*}
\caption[]{ OPTOPUS fields (available also in electronic form) }
\begin{flushleft}
\begin{tabular}{rrrrrrrr|rrrrrrrr}
\hline\noalign{\smallskip}
 \# & $\alpha$ & $\delta$ & $N_{T}$ & $N_{Z}$ & $N_{NO}$ &
 $N_{F}$ & $N_{S}$ & 
 \# & $\alpha$ & $\delta$ & $N_{T}$ & $N_{Z}$ & $N_{NO}$ &
 $N_{F}$ & $N_{S}$ \\ 
 & (1950) & (1950) & & & & & & & (1950) & (1950) & & & & & \\ 
\noalign{\smallskip}
\hline\noalign{\smallskip}
  ~1 &  22 31 18.3 &  -40 00 00 &  70 &  36 &  25 &   2 &   7 &
 101 &  22 31 18.3 &  -40 30 00 &  74 &  36 &  28 &   5 &   5 \\
  ~2 &  22 33 55.0 &  -40 00 00 &  56 &  33 &  11 &   8 &   4 & 
 102 &  22 33 55.0 &  -40 30 00 &  62 &  39 &  19 &   1 &   3 \\  
  ~3 &  22 36 31.6 &  -40 00 00 &  72 &  33 &  30 &   5 &   4 & 
 103 &  22 36 31.6 &  -40 30 00 &  42 &  34 &   3 &   1 &   4 \\  
  ~4 &  22 39 08.3 &  -40 00 00 &  77 &  28 &  34 &   9 &   6 &
 104 &  22 39 08.3 &  -40 30 00 &  36 &  30 &   2 &   1 &   3 \\  
  ~5 &  22 41 44.9 &  -40 00 00 &  80 &  41 &  32 &   1 &   6 &
 105 &  22 41 44.9 &  -40 30 00 &  57 &  34 &  12 &   4 &   7 \\  
  ~6 &  22 44 21.6 &  -40 00 00 &  53 &  32 &  10 &   1 &  10 &
 106 &  22 44 21.6 &  -40 30 00 &  53 &  23 &  16 &   8 &   6 \\  
  ~7 &  22 46 58.2 &  -40 00 00 &  69 &  34 &  25 &   1 &   9 &
 107 &  22 46 58.2 &  -40 30 00 &  92 &  37 &  43 &   0 &  12 \\  
  ~8 &  22 49 34.9 &  -40 00 00 &  49 &  37 &   4 &   3 &   5 &
 108 &  22 49 34.9 &  -40 30 00 &  54 &  31 &   9 &   2 &  12 \\  
  ~9 &  22 52 11.5 &  -40 00 00 &  43 &  25 &   1 &   3 &  14 &
 109 &  22 52 11.5 &  -40 30 00 &  57 &  30 &  13 &   6 &   8 \\  
  21 &  23 23 31.3 &  -40 00 00 &  27 &  19 &   0 &   0 &   8 &
 121 &  23 23 31.3 &  -40 30 00 &  55 &  37 &   5 &   2 &  11 \\  
  22 &  23 26 08.0 &  -40 00 00 &  43 &  33 &   1 &   0 &   9 & 
 122 &  23 26 08.0 &  -40 30 00 &  42 &  32 &   0 &   3 &   7 \\  
  23 &  23 28 44.6 &  -40 00 00 &  32 &  30 &   0 &   2 &   0 & 
 123 &  23 28 44.6 &  -40 30 00 &  25 &  21 &   0 &   0 &   4 \\ 
  24 &  23 31 21.2 &  -40 00 00 &  36 &  28 &   1 &   5 &   2 & 
 124 &  23 31 21.2 &  -40 30 00 &  42 &  36 &   2 &   3 &   1 \\  
  25 &  23 33 57.9 &  -40 00 00 &  48 &  37 &   1 &   2 &   8 & 
 125 &  23 33 57.9 &  -40 30 00 &  39 &  34 &   0 &   0 &   5 \\ 
  26 &  23 36 34.5 &  -40 00 00 &  36 &  26 &   2 &   4 &   4 & 
 126 &  23 36 34.5 &  -40 30 00 &  31 &  24 &   1 &   1 &   5 \\  
  27 &  23 39 11.2 &  -40 00 00 &  24 &   9 &   1 &   6 &   8 & 
 127 &  23 39 11.2 &  -40 30 00 &  35 &  28 &   1 &   0 &   6 \\  
  28 &  23 41 47.8 &  -40 00 00 &  22 &  17 &   1 &   2 &   2 & 
 128 &  23 41 47.8 &  -40 30 00 &  29 &  21 &   0 &   0 &   8 \\ 
  29 &  23 44 24.5 &  -40 00 00 &  24 &  23 &   0 &   0 &   1 &
 129 &  23 44 24.5 &  -40 30 00 &  15 &  13 &   0 &   0 &   2 \\ 
  30 &  23 47 01.1 &  -40 00 00 &  19 &  16 &   0 &   1 &   2 & 
 130 &  23 47 01.1 &  -40 30 00 &  23 &  19 &   1 &   0 &   3 \\  
  31 &  23 49 37.8 &  -40 00 00 &  32 &  28 &   1 &   0 &   3 & 
 131 &  23 49 37.8 &  -40 30 00 &  36 &  33 &   1 &   0 &   2 \\  
  32 &  23 52 14.4 &  -40 00 00 &  48 &  43 &   0 &   0 &   5 & 
 132 &  23 52 14.4 &  -40 30 00 &  22 &  19 &   0 &   0 &   3 \\ 
  33 &  23 54 51.1 &  -40 00 00 &  29 &  21 &   1 &   2 &   5 & 
 133 &  23 54 51.1 &  -40 30 00 &  42 &  35 &   2 &   1 &   4 \\  
  34 &  23 57 27.7 &  -40 00 00 &  56 &  34 &   9 &   1 &  12 & 
 134 &  23 57 27.7 &  -40 30 00 &  32 &  18 &   2 &   0 &  12 \\  
  35 &  00 00 04.4 &  -40 00 00 &  38 &  33 &   0 &   1 &   4 & 
 135 &  00 00 04.4 &  -40 30 00 &  34 &  27 &   0 &   0 &   7 \\ 
  36 &  00 02 41.0 &  -40 00 00 &  38 &  24 &   1 &   1 &  12 & 
 136 &  00 02 41.0 &  -40 30 00 &  53 &  42 &   4 &   2 &   5 \\  
  37 &  00 05 17.7 &  -40 00 00 &  39 &  33 &   2 &   1 &   3 & 
 137 &  00 05 17.7 &  -40 30 00 &  32 &  25 &   2 &   2 &   3 \\  
  38 &  00 07 54.3 &  -40 00 00 &  29 &  26 &   1 &   2 &   0 & 
 138 &  00 07 54.3 &  -40 30 00 &  18 &  15 &   1 &   0 &   2 \\  
  39 &  00 10 31.0 &  -40 00 00 &  37 &  25 &   2 &   3 &   7 & 
 139 &  00 10 31.0 &  -40 30 00 &  25 &  20 &   1 &   3 &   1 \\  
  40 &  00 13 07.6 &  -40 00 00 &  53 &  46 &   1 &   4 &   2 & 
 140 &  00 13 07.6 &  -40 30 00 &  46 &  35 &   6 &   4 &   1 \\  
  41 &  00 15 44.3 &  -40 00 00 &  47 &  38 &   2 &   4 &   3 & 
 141 &  00 15 44.3 &  -40 30 00 &  39 &  33 &   3 &   2 &   1 \\  
  42 &  00 18 20.9 &  -40 00 00 &  36 &  29 &   5 &   1 &   1 & 
 142 &  00 18 20.9 &  -40 30 00 &  33 &  30 &   0 &   0 &   3 \\ 
  43 &  00 20 57.6 &  -40 00 00 &  33 &  22 &   0 &  11 &   0 & 
 143 &  00 20 57.6 &  -40 30 00 &  68 &  63 &   2 &   0 &   3 \\ 
  44 &  00 23 34.2 &  -40 00 00 &  49 &  40 &   6 &   0 &   3 & 
 144 &  00 23 34.2 &  -40 30 00 &  43 &  38 &   4 &   0 &   1 \\ 
  45 &  00 26 10.9 &  -40 00 00 &  40 &  29 &   2 &   7 &   2 & 
 145 &  00 26 10.9 &  -40 30 00 &  25 &  25 &   0 &   0 &   0 \\ 
  46 &  00 28 47.5 &  -40 00 00 &  24 &  15 &   0 &   7 &   2 & 
 146 &  00 28 47.5 &  -40 30 00 &  27 &  22 &   2 &   0 &   3 \\  
  47 &  00 31 24.2 &  -40 00 00 &  23 &  18 &   1 &   2 &   2 & 
 147 &  00 31 24.2 &  -40 30 00 &  25 &  20 &   1 &   1 &   3 \\  
  48 &  00 34 00.8 &  -40 00 00 &  36 &  29 &   1 &   4 &   2 & 
 148 &  00 34 00.8 &  -40 30 00 &  31 &  23 &   1 &   2 &   5 \\  
  49 &  00 36 37.5 &  -40 00 00 &  28 &  22 &   0 &   0 &   6 & 
 149 &  00 36 37.5 &  -40 30 00 &  21 &  18 &   1 &   0 &   2 \\  
  50 &  00 39 14.1 &  -40 00 00 &  24 &  21 &   0 &   2 &   1 & 
 150 &  00 39 14.1 &  -40 30 00 &  20 &  18 &   0 &   0 &   2 \\ 
  51 &  00 41 50.8 &  -40 00 00 &  54 &  44 &   2 &   2 &   6 & 
 151 &  00 41 50.8 &  -40 30 00 &  30 &  28 &   0 &   0 &   2 \\ 
  52 &  00 44 27.4 &  -40 00 00 &  44 &  37 &   1 &   1 &   5 & 
 152 &  00 44 27.4 &  -40 30 00 &  42 &  35 &   0 &   2 &   5 \\  
  53 &  00 47 04.1 &  -40 00 00 &  59 &  58 &   0 &   0 &   1 & 
 153 &  00 47 04.1 &  -40 30 00 &  43 &  37 &   3 &   1 &   2 \\  
  54 &  00 49 40.7 &  -40 00 00 &  48 &  42 &   1 &   1 &   4 & 
 154 &  00 49 40.7 &  -40 30 00 &  38 &  31 &   1 &   3 &   3 \\  
  55 &  00 52 17.4 &  -40 00 00 &  61 &  54 &   1 &   4 &   2 & 
 155 &  00 52 17.4 &  -40 30 00 &  37 &  24 &   2 &   1 &  10 \\  
  56 &  00 54 54.0 &  -40 00 00 &  39 &  31 &   1 &   1 &   6 & 
 156 &  00 54 54.0 &  -40 30 00 &  40 &  36 &   0 &   2 &   2 \\  
  57 &  00 57 30.7 &  -40 00 00 &  60 &  56 &   0 &   0 &   4 &
 157 &  00 57 30.7 &  -40 30 00 &  67 &  45 &   6 &   3 &  13 \\   
  58 &  01 00 07.3 &  -40 00 00 &  47 &  42 &   0 &   2 &   3 & 
 158 &  01 00 07.3 &  -40 30 00 &  48 &  40 &   3 &   0 &   5 \\   
  59 &  01 02 44.0 &  -40 00 00 &  83 &  78 &   0 &   0 &   5 &
 159 &  01 02 44.0 &  -40 30 00 &  59 &  48 &   5 &   2 &   4 \\   
  60 &  01 05 20.6 &  -40 00 00 &  40 &  33 &   2 &   0 &   5 & 
 160 &  01 05 20.6 &  -40 30 00 &  74 &  65 &   4 &   2 &   3 \\   
  61 &  01 07 57.3 &  -40 00 00 &  33 &  21 &   1 &   4 &   7 & 
 161 &  01 07 57.3 &  -40 30 00 &  43 &  31 &   0 &   6 &   6 \\   
  62 &  01 10 33.9 &  -40 00 00 &  28 &  23 &   1 &   0 &   4 & 
 162 &  01 10 33.9 &  -40 30 00 &  34 &  31 &   1 &   0 &   2 \\   
  63 &  01 13 10.6 &  -40 00 00 &  31 &  26 &   0 &   2 &   3 & 
 163 &  01 13 10.6 &  -40 30 00 &  34 &  27 &   1 &   1 &   5 \\   
  64 &  01 15 47.2 &  -40 00 00 &  26 &  19 &   1 &   2 &   4 & 
 164 &  01 15 47.2 &  -40 30 00 &  36 &  29 &   1 &   1 &   5 \\   
  65 &  01 18 23.9 &  -40 00 00 &  55 &  41 &   4 &   2 &   8 & 
     &             &            &     &     &     &     &     \\
\noalign{\smallskip}
\hline
\end{tabular}
\end{flushleft}
\end{table*}

\section{Catalogue presentation}

In Table 2 we list the fields we observed with OPTOPUS (see Sect.1 and Paper I)
and their properties. Column (1) gives the number of the field, whose centre
coordinates ($\alpha$ and $\delta$) are reported in columns (2) and (3).
Column (4) gives the total number of objects classified as galaxies
in the photometric catalogue ($N_T$), while
columns (5), (6), (7) and (8) give the number of redshifts ($N_Z$), of 
not--observed objects ($N_{NO}$), of failed spectra (i.e. not useful to obtain 
a redshift determination, $N_F$) and of stars ($N_S$), respectively. 
From these numbers the redshift completeness of each field can be derived
(see also Paper I and Zucca et al. 1997) as

\begin{equation}
{ { N_Z } \over { N_T - N_S - 0.122 * N_{NO} } }
\end{equation}

This equation assumes that the failed spectra 
correspond to galaxies, because stars, being point--like objects, have
on average a better signal--to--noise ratio than galaxies, and 
that the percentage of stars in not--observed objects spectra is the same as
in the spectroscopic sample (i.e. $\sim 12.2\%$). 

Note that the field centres are separated by 30 arcmin both in right ascension
and declination whilst the OPTOPUS fields have a diameter of 32 arcmin.
Therefore there is a small overlap between the fields which results in the fact
that some galaxies belong to two adjacent fields; the galaxies in the overlap 
areas have been assigned to the field whose centre is closer to the
object position. 

\begin{table*}
\caption[]{ Sample page of the catalogue (the whole catalogue is available 
only in electronic form) }
\begin{flushleft}
\begin{tabular}{rrrrrrrrrr}
\hline\noalign{\smallskip}
 \# object & \# field & $\alpha$ & $\delta$ & $b_J$ & 
$v_{abs}$ & $err$ & R & $v_{emiss}$ & $err$ \\
 ~~~~~~~~~ & ~~~~~~~~ &   (1950) &   (1950) & ~~~~  & 
(km/s) & (km/s) & ~ & (km/s) & (km/s) \\
\noalign{\smallskip}
\hline\noalign{\smallskip}
 32986 & 147 & 00 31 08.5 & -40 38 42 & 19.32 & 56575& 106&  2.34&  56509&  60\\
 33018 & 147 & 00 31 12.8 & -40 43 28 & 19.15 &     0&   0&  0.00&  37564&  14\\
 34795 & 147 & 00 31 16.4 & -40 30 32 & 18.70 & 35523&  47&  7.42&      0&   0\\
 34801 & 147 & 00 31 21.5 & -40 23 11 & 18.37 & -9999&   0&  0.00&  -9999&   0\\
 34800 &  47 & 00 31 22.0 & -39 46 42 & 18.76 & 32884&  37&  9.80&      0&   0\\
 34802 & 147 & 00 31 30.3 & -40 33 12 & 17.94 & 33643&  30& 11.30&      0&   0\\
 33049 & 147 & 00 31 33.9 & -40 35 05 & 19.19 & 33875&  74&  4.31&  33775&  71\\
 34805 &  47 & 00 31 39.0 & -40 03 12 & 18.39 & 28517&  42&  9.81&      0&   0\\
 33089 & 147 & 00 31 45.7 & -40 18 47 & 19.39 &     0&   0&  0.00&  20619&  20\\
 34810 & 147 & 00 31 46.5 & -40 17 19 & 17.98 & 19991&  33& 12.69&      0&   0\\
 34811 & 147 & 00 31 48.3 & -40 17 40 & 18.71 & 20177&  20& 11.66&      0&   0\\
 33091 & 147 & 00 31 51.3 & -40 31 53 & 19.25 & 34073& 103&  3.48&  33945&  41\\
 34820 &  47 & 00 31 54.9 & -39 50 46 & 18.72 & 33519&  34& 11.63&      0&   0\\
 33132 &  47 & 00 32 09.9 & -40 08 53 & 18.96 & 28486&  54&  4.18&  28637&  37\\
 33130 &  47 & 00 32 12.0 & -40 01 54 & 18.89 & 32838&  49&  6.50&  32844&  37\\
 34826 & 147 & 00 32 15.0 & -40 32 09 & 17.65 & 20032&  64&  5.51&      0&   0\\
 33163 &  47 & 00 32 19.8 & -39 59 56 & 18.98 &     0&   0&  0.00&  26223&  23\\
 34831 &  47 & 00 32 20.3 & -39 50 58 & 18.61 & -8888&   0&  0.00&  -8888&   0\\
 33161 &  47 & 00 32 21.4 & -39 56 27 & 19.00 & 54961& 117&  1.70&      0&   0\\
 33168 & 147 & 00 32 25.7 & -40 19 32 & 18.92 & 20073&  88&  4.10&  20115&  42\\
 33164 &  47 & 00 32 27.2 & -40 01 58 & 19.34 & 73433&  99&  3.18&      0&   0\\
 33162 &  47 & 00 32 27.3 & -39 57 16 & 18.95 & 55121&  56&  6.91&      0&   0\\
 34832 &  47 & 00 32 27.6 & -40 06 17 & 18.63 & 33059&  91&  3.22&      0&   0\\
 33165 &  47 & 00 32 28.3 & -40 04 11 & 19.32 &     0&   0&  0.00&      0&   0\\
 33166 &  47 & 00 32 28.9 & -40 04 25 & 18.93 &     0&   0&  0.00&  11749&  14\\
 33169 & 147 & 00 32 32.8 & -40 22 06 & 18.81 &     0&   0&  0.00&  20186&  67\\
 33189 &  47 & 00 32 33.4 & -39 51 26 & 18.87 & -8888&   0&  0.00&  -8888&   0\\
 34840 & 147 & 00 32 34.1 & -40 28 10 & 18.59 & 24768&  59&  5.45&      0&   0\\
 33220 &  48 & 00 32 44.9 & -40 03 27 & 19.32 & 73218&  68&  4.00&      0&   0\\
 34848 & 148 & 00 32 47.0 & -40 23 12 & 18.42 & -9999&   0&  0.00&  -9999&   0\\
 33225 & 148 & 00 32 53.3 & -40 31 20 & 19.22 & 34012&  98&  4.20&      0&   0\\
 33223 & 148 & 00 32 53.8 & -40 25 04 & 18.97 & 61621&  82&  3.00&  61736&  55\\
 34849 & 148 & 00 32 54.5 & -40 30 05 & 18.65 & 34230&  94&  3.50&  34191&  37\\
 34850 & 148 & 00 32 57.4 & -40 32 54 & 18.41 &     0&   0&  0.00&      0&   0\\
\noalign{\smallskip}
\hline
\end{tabular}
\end{flushleft}
\end{table*}

We observed a total of 4043 objects, corresponding to $\sim 90\%$ of the 
parent photometric sample of 4487 objects.
Out of the 4043 observed objects, 493 turned out to be stars and 
207 have a too low signal--to--noise ratio to provide a reliable redshift
(failed spectra). In the end, our final sample consists of a total of 
3343 objects with reliable redshifts (3342 galaxies and 1 QSO). 

The survey data are available in electronic form at the CDS via anonymous ftp 
to cdsarc.u-strasbg.fr (130.79.128.5) or via 
http://cdsweb.u-strasbg.fr/Abstract.html. 
In Table 3 we provide a sample page of the catalogue, which is sorted in right 
ascension. The columns contain the following information:
\\
Column (1): ESP galaxy number
\\
Column (2): OPTOPUS field number
\\
Column (3): Right Ascension (1950)
\\
Column (4): Declination (1950)
\\
Column (5): $b_J$ magnitude
\\
Column (6): Heliocentric Radial Velocity from absorption lines in km/s
\\
Column (7): Associated Internal Error in km/s
\\
Column (8): Value of the R parameter (Tonry \& Davis 1979) from cross 
correlation 
\\
Column (9): Heliocentric Radial Velocity from emission lines in km/s
\\
Column (10): Associated Error in km/s

The codes $-9999$ and $-8888$ in the velocity columns indicate stellar
spectra (hence stars misclassified in the EDSGC) and spectra not useful
for radial velocity measurements, respectively. 
Objects which have not been observed have a zero in the velocity columns. 
For two galaxies (\# 13489 = NGC 7410 and \# 17863) the reported velocities 
are from da Costa et al. (1991) and Metcalfe et al. (1989), respectively. 
The object \# 31954, with the velocity coded as $99999$, is a quasar with 
$z \sim 1.174$.
Finally, a few galaxies have a measure for $v_{abs}$ but have $R=0$: this fact
indicates low quality spectra for which the cross--correlation was ``forced",
choosing the correlation peak by hand.

In the case of multiple observations of the same galaxy, we report
in Table 3 the best measurement only.

For what concerns the use of these data for scientific analyses, 
when a galaxy has both $v_{abs}$ and $v_{emiss}$ the choice of the velocity can
be done on the basis of the minimum error: however, the differences between 
$v_{abs}$ and $v_{emiss}$ are so small that different choices do not produce
appreciable effects on the results of most of the scientific analyses.
Note that the velocity errors reported in column (7) and (10) are formal 
errors: for the conversion factors from these internal errors to the true 
errors, see the discussion in Section 6.2.

\begin{figure}
\epsfysize=9cm
\epsfbox{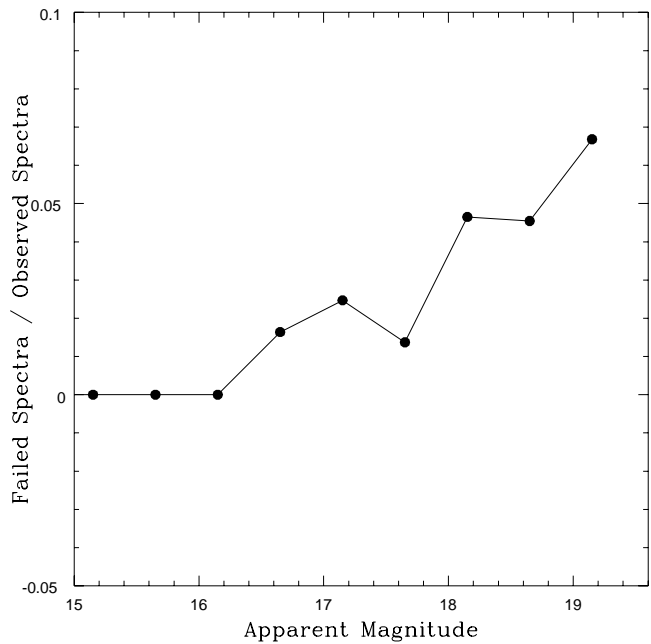}
\caption[]{Fraction of failed spectra over the total number of observed objects
as a function of magnitude. Note that, even in the faintest bin, this fraction
is always lower than $7\%$. }
\end{figure}

\begin{figure}
\epsfysize=9cm
\epsfbox{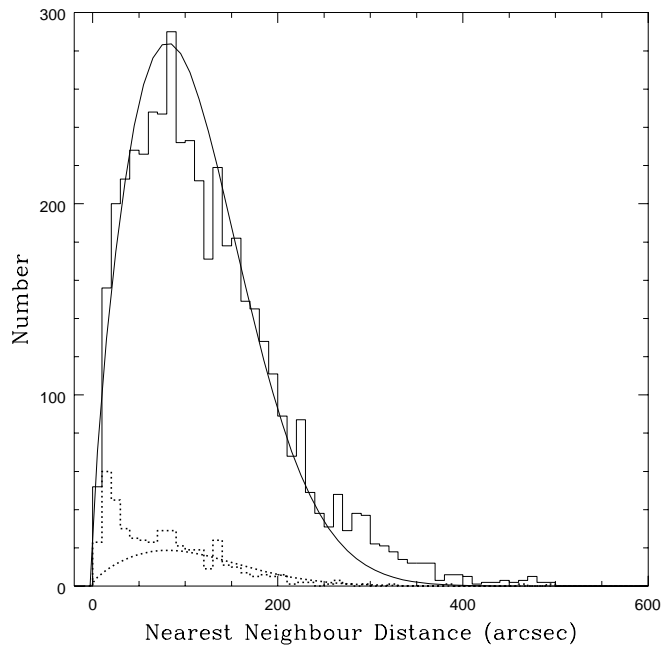}
\caption[]{ Observed distribution of nearest neighbour distances ($d_{nn}$) 
for the entire photometric catalogue (4487 objects; solid histogram) and for 
the 444 objects which have not been observed spectroscopically (dashed 
histogram). The curves superimposed on the two histograms show the expected 
distributions of $d_{nn}$ on the basis of the measured angular correlation 
function. }
\end{figure}

\section{Analysis}

\subsection{ Possible biases in the sample } 

Statistical analyses of this sample have to take into account and, if
necessary, to correct for possible biases induced by the fact that some
stars are misclassified as galaxies in the photometric catalogue, not all
spectra have produced a measurable redshift and not all objects have been
observed.

We have verified that the magnitude distribution of the stars is consistent
with that of the total sample and therefore does not bias any analyses. 

The small fraction ($\sim 5\%$) of objects with unmeasurable redshift
has various origins: $\sim 13 \%$ of them correspond to undeblended pairs (see
Sect.2) for which the object coordinates fall in between the two components,
while most cases are due to not well connected fibers or to observations with 
bad weather conditions. This kind of incompleteness is
higher for fainter objects (see Figure 2): note, however, that the 
maximum fraction of galaxies 
for which the spectra did not provide a useful $z$ determination is $\sim 7\%$ 
for the faintest galaxies of our survey ($b_J = 19.4$).

We have verified that the magnitude distribution of not--observed objects 
($\sim 10\%$) is consistent with being a random extraction from the total 
sample. The main bias for such objects is introduced
by the impossibility of observing, in a given OPTOPUS exposure, two
objects closer than $\sim$ 25 arcsec. Although the MEFOS observations
and the repeated observations of the same fields have reduced somewhat
this bias, it is still significantly present in the data. This is
clearly seen in Figure 3 which shows the observed distribution of
nearest neighbour distances ($d_{nn}$) for the entire photometric catalogue 
(4487 objects; solid histogram) and for the 444 objects which
have not been observed spectroscopically (dashed histogram). The curves
superimposed on the two histograms show the expected distributions of
$d_{nn}$ on the basis of the measured angular correlation function. The
excess of not--observed objects which have a neighbour at a distance
smaller than 50 arcsec is 132 $\pm$ 14 (183 objects in the data while 54 would
be expected); some, less significant excess (35 $\pm$ 11) is present also for 
$d_{nn}$ in the range 50--100 arcsec. Another way of characterizing the 
same bias is through the ratio between the not--observed objects and the
total number of objects in the photometric catalogue as a function of 
$d_{nn}$. This ratio is 0.31, 0.13 0.10, 0.06 for $d_{nn} < 30$, 
$30 \le d_{nn} < 50$, $50 \le d_{nn} < 100$, $100 \le d_{nn} $, 
respectively. Because of this bias, the spectroscopic catalogue can not
be used in a straightforward way for studying, for example, the statistics
of the number of close pairs or for analyzing the three dimensional correlation
function on very small scales.

\subsection{Statistical analysis of velocity errors} 

The errors on the velocities listed in Columns (7) and (10) of Table 3 are
the formal errors given by the {\sc iraf} tasks {\sc xcsao} and {\sc emsao} 
for $v_{abs}$ and $v_{emiss}$, respectively. Their distributions are shown in
Figure 4: the median error is 64 km/s for $v_{abs}$ and 31 km/s for $v_{emiss}$.
In order to have a better
estimate of the true errors, we have analyzed the differences in the
measured velocities ($\Delta v$) of the galaxies which have been observed
more than once (156 galaxies with two measurements of $v_{abs}$ and 64 
galaxies with two measurements of $v_{emiss}$).
For both these samples the histograms of $\Delta v$ normalized to the formal
errors are significantly larger than expected if these formal errors were 
correct estimates of the true errors. 
By fitting these distributions with Gaussian
curves, the dispersions of the best fitting Gaussians are 1.53 and 2.10
for $v_{abs}$ and $v_{emiss}$, respectively. This implies that, assuming that
the ratio between the true and the formal errors is a constant for all
measurements, the true errors can be estimated by multiplying the errors
given in Table 3 by these factors.  

\begin{figure}
\epsfysize=9cm
\epsfbox{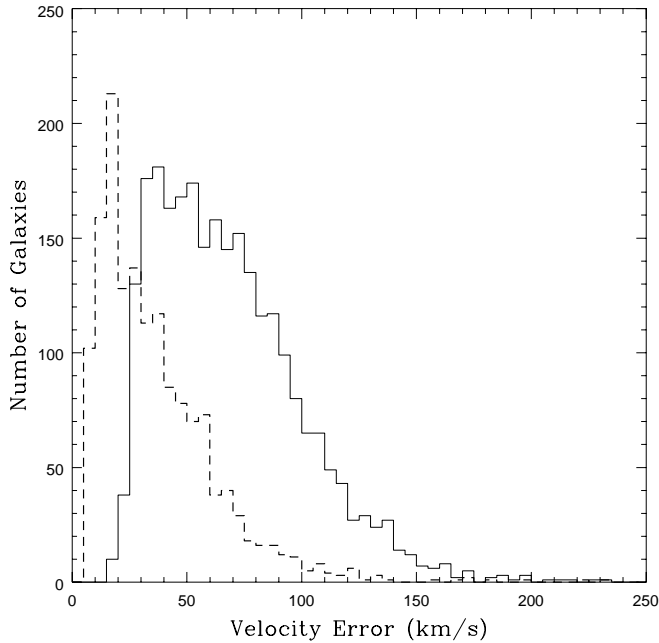}
\caption[]{ Distribution of the errors on the measured velocity, for both
$v_{abs}$ (solid histogram) and $v_{emiss}$ (dashed histogram). }
\end{figure}

The factor found for $v_{abs}$ is in agreement with other values reported in 
the literature from similar analyses: for instance, Bardelli et al. (1994) 
found a factor 1.87 from a comparison of two sets of OPTOPUS observations
(45 galaxies) reduced with different packages ({\sc iraf} and {\sc midas})
and by different authors; Malumuth et al. (1992), using multiple observations 
of 42 galaxies reduced in the same way, found a factor 1.6.

\subsection{Systematic difference between $v_{emiss}$ and $v_{abs}$}

For about 750 galaxies we could measure the redshift both
from absorption and from emission lines. The distribution of the difference 
($v_{abs} - v_{emiss}$) is well fitted by a Gaussian
peaked at $\sim 100$ km/s.
We have performed a number of tests, and we can exclude a zero--point
error. Also in the Las Campanas redshift survey a similar systematic difference 
between the absorption and the emission velocities 
has been found, and the authors have chosen to correct for it by using a 
separate average template with large Balmer lines to fit emission line
galaxies (Shectman et al. 1996). 
They suggest that the systematic effect is mainly due to the
blend between $H\epsilon$ and $Ca H$ lines.
On the other hand, we have decided to use in any case
the best--fitting template, for sake of homogeneity in the 
redshift measures.
A more general discussion about this point will be found in Cappi et al.
(1998).

\section{Summary}

We have described in detail the data of the ESP galaxy redshift survey,
which extends over about 23 square degrees, in a region near the South
Galactic Pole. The survey is $\sim 85\%$ complete to the limiting magnitude
$b_J=19.4$ and consists of 3342 galaxies with redshift determination. 
Although not all galaxies have been observed and not all spectra
have produced a measurable redshift, we have shown that these facts
do not introduce any bias in the final spectroscopic
sample. The only significant bias still remaining in the sample
is due to the fact that close pairs of galaxies could not
be observed in a single OPTOPUS (or MEFOS) observation. For this reason
the fraction of not--observed objects is significantly higher than average
for objects which have a companion in the photometric catalogue at a distance
smaller than about 50 arcsec.
\\
For all galaxies we have determined, when possible, both absorption
and emission velocities. The median formal errors on the velocities
are 64 and 31 km/s for the absorption and emission velocities, respectively.
Analysis of the velocity measurements of the galaxies which have been observed
more than once shows, however, that these formal errors are significant
underestimates of the ``true'' errors. In first approximation the true
errors can be obtained by multiplying the formal ones by factors of the
order of 1.5 and 2.1 for $v_{abs}$ and $v_{emiss}$, respectively.
\\
The data of the catalogue, available in electronic form at the  
the CDS via anonymous ftp to cdsarc.u-strasbg.fr 
(130.79.128.5) or via http://cdsweb.u-strasbg.fr/Abstract.html,
provide all the information which is needed
(i.e. positions, magnitudes, velocities, completeness) for statistical
analyses of this sample, as for example the estimate of the luminosity
function and mean galaxy density (Zucca et al. 1997).


\begin{acknowledgements}

This work has been partially supported through NATO Grant CRG 920150,  
EEC Contract ERB--CHRX--CT92--0033, CNR Contract 95.01099.CT02 and by 
Institut National des Sciences de l'Univers and Cosmology GDR.
\\
It is a pleasure to thank the support we had from the ESO staff both in 
La Silla and in Garching. In particular, we are grateful to Gerardo Avila for 
his advice and his help in solving every instrumental problem we have been
facing during this project. 
We also thank M. Kurtz and D. Mink who kindly provided the
{\sc rvsao} package before the official release in {\sc iraf}.

\end{acknowledgements}



\end{document}